\documentclass[aps,nofootinbib,twocolumn,floatfix,prd,letterpaper,superscriptaddress]{revtex4}
\usepackage{graphicx,amsmath,amssymb,color,comment,subfigure}

\def\be{\begin{equation}}
\def\ee{\end{equation}}
\def\bea{\begin{eqnarray}}
\def\eea{\end{eqnarray}}
\newcommand{\vs}{\nonumber\\}
\def\ba#1\ea{\begin{align}#1\end{align}}
\def\bg#1\eg{\begin{gather}#1\end{gather}}

\newcommand{\fsky}{f_{\rm sky}}

\newcommand{\g}{\gamma}

\newcommand{\refeq}[1]{Eq.~(\ref{eq:#1})}

\renewcommand{\v}[1]{\mathbf{#1}}

\newcommand{\vx}{\v{x}}

\newcommand{\vk}{\v{k}}

\newcommand{\<}{\langle}
\renewcommand{\>}{\rangle}

\renewcommand{\d}{\delta}

\newcommand{\Om}{\Omega_m}

\def\L{\mathcal{L}}

\renewcommand{\Im}{{\rm Im}\,}

\begin{document}

\title{Can weak lensing surveys confirm BICEP2 ?}

\author{Nora Elisa Chisari}
\altaffiliation{nchisari@astro.princeton.edu} 
\affiliation{Department of Astrophysical Sciences, Princeton University, 4 Ivy Lane, Princeton, NJ~08544, USA}

\author{Cora Dvorkin}
\affiliation{Institute for Advanced Study, School of Natural Sciences, Einstein Drive, Princeton, NJ~08540, USA}

\author{Fabian Schmidt}
\affiliation{Max-Planck-Insitute for Astrophysics, D-85748 Garching, Germany}

\begin{abstract}
The detection of B-modes in the Cosmic Microwave Background (CMB) polarization by the BICEP2 experiment, if interpreted as evidence for a primordial gravitational wave background, has enormous ramifications for cosmology and physics. It is crucial to test this hypothesis with independent measurements. A gravitational wave background leads to B-modes in galaxy shape correlations (shear) both through lensing and tidal alignment effects. Since the systematics and foregrounds of galaxy shapes and CMB polarization are entirely different, a detection of a cross-correlation between the two observables would provide conclusive proof for the existence of a primordial gravitational wave background. We find that upcoming weak lensing surveys will be able to detect the cross-correlation between B-modes of the CMB and galaxy shapes. However, this detection is not sufficient to confirm or falsify the hypothesis of a primordial origin for CMB B-mode polarization. 
\end{abstract}

\date{\today}
\maketitle

%%%%%%%%%%%%%%%%%%%%%%%%%%%%%%%%%%%%%%%%%%%%%%%%%%%%%%%%%%%%%%%%%%%%%%%%%%%
%%%%%%%%%%%%%%%%%%%%%%%%%%%%%%%%%%%%%%%%%%%%%%%%%%%%%%%%%%%%%%%%%%%%%%%%%%%
\section{Introduction}
\label{sec:intro}

In the foreseeable future, galaxy shape correlations (shear) are perhaps the most promising way of confirming the possible detection of gravitational waves in the B-mode polarization of the Cosmic Microwave Background (CMB). 

Gravitational waves are sourced by quantum fluctuations that are exponentially stretched during inflation. Their amplitude is frozen as they exit the horizon. Upon re-entry, these waves create anisotropies in the density of free electrons. Thomson scattering of the CMB radiation impinging on free electrons before recombination ($z\simeq1100$), and during reionization (at $z\simeq 10$) produces a B-mode polarization signal \cite{Seljak97,Kamionkowski97,Krauss10}. Constraints on the amplitude of the power spectrum of tensor modes can be obtained from measuring the CMB B-mode pattern \citep{Komatsu11,Bicep2a}. However, a similar B-mode pattern can also be produced by foregrounds in our Galaxy \cite{Flauger14,Mortonson/Seljak}. Thus, the exploration of other probes of gravitational waves from inflation, subject to different systematics, is a promising avenue for confirmation of their detection.

The shapes of galaxies are distorted by the effect of tidal fields of the large-scale structure (``intrinsic alignments'', \citet{Catelan01}) and by gravitational lensing due to the spacetime perturbations along the line of sight \citep{Kaiser92}. Tensor modes contribute to both effects \citep{Schmidt12,Schmidt14}. They induce an effective tidal field that correlates the shape of a galaxy with the time derivative of tensor modes after they enter the horizon, leading to shear B-modes. They also contribute to lensing by deflecting light along the path to the observer \cite{kaiser/jaffe,dodelson/rozo/stebbins,Schmidt12}.  Another promising probe of primordial gravitational waves is offered by the curl-component of the lensing of fluctuations in the 21 cm emission from the dark ages \cite{MasuiPen,Book12}.  However, 21 cm surveys capable of reaching the required sensitivity level remain many decades in the future.  Inhomogeneous reionization can also be employed to search for tensor modes in cross-correlation with the CMB polarization \cite{alizadeh/hirata}.  

In this work, we explore the prospects of detecting the cross-correlation of galaxy shapes with CMB polarization B-modes from gravitational waves with upcoming surveys. The systematic issues affecting the two measurements are entirely independent. As a consequence, the cross-correlation is more robust to systematics, allowing for a possible confirmation. First, these measurements employ completely different measurement techniques, radio-band polarization measurements on tens of arcmin scales and above on the one hand, and arcsecond-resolution optical imaging on the other.  Second, the main foregrounds of the CMB measurements, polarized dust and synchrotron emission from the Galaxy, do not affect the measurements of galaxy shapes in the optical.  Conversely, the second-order scalar contribution to shear correlations (from lensing bias and reduced shear in particular) contaminate the tensor mode signal in the shear, but do not contribute to CMB polarization on degree scales and larger. Thus, while the auto-correlation of shapes or B-modes remain subject to systematics, the cross-correlation between both probes could provide an exceptionally clean and convincing confirmation that the B-mode signal detected is indeed of primordial origin. 

\citet{Dodelson10} studied the cross-correlation of the galaxy shear produced by weak gravitational lensing and CMB B-modes from reionization.  There are two main differences between our treatment and that of Ref. \cite{Dodelson10}: first, we include the contribution of intrinsic alignments to the shear, which have been shown to dominate the tensor mode signal in shape correlations \cite{Schmidt12,Schmidt14}.  Second, we also use the full numerical computation of the B-mode polarization transfer function for tensors, which includes the contribution from reionization and recombination.

This work is organized as follows. In Section \ref{sec:IA}, we give the expressions for the auto-correlation of CMB B-modes from tensor modes, the shear B-mode auto-correlation and the cross-correlation of these observables. In Section \ref{sec:sn1}, we estimate the likelihood of detecting this cross-correlation and for ruling out the scenario where the CMB B-modes are due to foregrounds (such as Galactic dust) with an \emph{Euclid}-like survey\footnote{http://sci.esa.int/euclid/}. We discuss our results in Section \ref{sec:concl}.

Throughout, we will assume a scalar-to-tensor ratio of $r = 0.2$ at $k_0= 0.002\;{\rm Mpc}^{-1}$, corresponding to the best-fit value without foregruonds claimed by the BICEP2 experiment \cite{Bicep2a}. \emph{Planck} on the other hand has obtained a bound on $r<0.12$ at the 95\% confidence level \cite{Planck13}. The discrepancy between these experiments is as yet unresolved, but could be due to Galactic foregrounds \cite{Flauger14,Mortonson/Seljak}.
Together with our fiducial cosmology, $r$ determines $\Delta_T^2$, the amplitude of the tensor modes power spectrum at $k_0$.  The tensor mode power spectrum is
\be
P_{T0}(k) = 2\pi^2\,k^{-3}\,\left(\frac{k}{k_0}\right)^{n_T}\Delta_T^2
\ee
The tensor index is chosen to follow the inflationary consistency relation, $n_T = -r/8 = -0.025$.  For the expansion history, we assume a flat $\Lambda$CDM cosmology with $h=0.72$ and $\Omega_m=0.28$. Contributions from scalar perturbations are evaluated using a spectral index of $n_s=0.958$ and power spectrum normalization at $z=0$ of $\sigma_8 = 0.810$.

%%%%%%%%%%%%%%%%%%%%%%%%%%%%%%%%%%%%%%%%%%%%%%%%%%%%%%%%%%%%%%%%%%%%%%%%%%%
%%%%%%%%%%%%%%%%%%%%%%%%%%%%%%%%%%%%%%%%%%%%%%%%%%%%%%%%%%%%%%%%%%%%%%%%%%%
\section{Galaxy shape correlations from tensor modes}
\label{sec:IA}

Let us denote CMB polarization with $P$ and shear with $\gamma$.  We can write the $B$-mode angular auto and cross power spectra as
\ba
C^{XY}(l) =\:& \frac1{2\pi} 
\int k^2 dk\: P_{T0}(k) F_l^X(k) F_l^Y(k)\,,\label{eq:Cl} 
\ea
where $X,Y \in \{P, \gamma\}$, and the power spectrum of primordial tensor modes is defined through
\ba
\< h_{ij}(\vk) h^{ij}(\vk') \> =\:& (2\pi)^3 \d_D(\vk-\vk') P_{T0}(k) \,. 
\label{eq:PT}
\ea
Here, $h_{ij}(\vk)$ denotes the transverse-traceless metric perturbation evaluated far outside the horizon $k \tau \ll 1$, where $\tau$ denotes
conformal time.   Note that all angular power spectra appearing throughout the paper will be odd-parity $B$-mode power spectra.  The filter function for the shear is given by
\ba
F_l^{\g}(k) \equiv\:& 
- \frac14 
\bigg[ T_T(k,\tau_0) \left(\Im \hat Q_1(x) \frac{j_l(x)}{x^2}\right)_{x=0} \vs
& \quad\quad+ T_T(k,\tilde\tau)\Im \hat Q_1(\tilde x) \frac{j_l(\tilde x)}{\tilde x^2}\bigg] \vs
 & + \int_0^{\chi(\tilde\tau)} \frac{d\chi}\chi
\Im\hat Q_2(x) \frac{j_l(x)}{x^2} T_T(k, \tau_0-\chi) \vs
& - \Om C_I\: \frac72 \frac{\alpha(k, \tilde\tau)}{a(\tilde\tau)} \Im \hat Q_1(\tilde x) \frac{j_l(\tilde x)}{\tilde x^2}\,,
\label{eq:Flgamma}
\ea
where $\hat Q_i(x)$ are derivative operators whose action on $j_l(x)/x^2$ are given explicitly in Eq.~(B16) of \cite{Schmidt12} and $T_T(k,\tau) = 3 j_1(k\tau)/(k\tau)$ is the tensor mode transfer function. $\tau_0$ and $\tilde\tau$ indicate the conformal times today and at the source epoch, respectively, and $x = k\chi$, $\tilde x = k\chi(\tilde\tau)$.  
$\chi$ denotes the comoving distance.  The first three terms in \refeq{Flgamma} correspond to the gravitational lensing by tensor modes,\footnote{This also includes the ``metric shear'' term of \cite{dodelson/rozo/stebbins} which \cite{Dodelson10} does not include; note that this term has to be there to ensure a proper gauge-invariant result \cite{Schmidt12}.} while the last term is the contribution of tidal alignments, parametrized by the linear alignment coefficient $C_I$. The tidal alignment model \cite{Catelan01} has been shown to provide a good description of the intrinsic alignments of red galaxies on linear scales \cite{Blazek11}, which are the focus of our work. Blue galaxies, on the contrary, have currently no measured shape alignments \cite{Heymans13}. For a review of intrinsic alignments of galaxies, see \cite{Troxel14}.

As shown in \cite{Schmidt12,Schmidt14}, 
long-wavelength tensor modes locally induce a tidal field once they reenter the horizon.  This tidal field oscillates and decays as the tensor mode redshifts.  Nevertheless, it leads to a lasting imprint in the small-scale density field \cite{Schmidt14,MasuiPen} similar to the effect of a large-scale scalar tidal field (produced by density fluctuations) on small-scale density fluctuations.  It is well known that dark matter halos and galaxies align with large-scale scalar tidal fields. \citet{Schmidt14} estimated the corresponding alignment with tensor-mode induced tidal fields by matching the effect on small-scale density fluctuations which they calculated quantitatively using second-order perturbation theory.  Specifically, they derived the contribution to the second-order matter density
perturbation $\d_{2,t}$ induced by a tensor mode with primordial amplitude $h_{ij}^{(0)}$ and wavenumber $k$, leading to 
\ba
& \d_{2,t}(\vx,\tau) \label{eq:d2tintro}\\
& \quad = h_{ij}^{(0)}(\vx) \left[ \alpha(k,\tau) \frac{\partial^i\partial^j}{\nabla^2} + \beta(k,\tau) x^i \partial^j \right] \d_{\rm lin}(\vx,\tau)\,,
\nonumber
\ea
where $\d_{\rm lin}$ denotes the linear density field.  The coefficient function
$\alpha(k,\tau)$ appearing in \refeq{Flgamma} and \refeq{d2tintro} thus quantifies the coupling of the tensor and scalar tidal fields,
while $\beta$ corresponds to a differential displacement effect which we 
ignore here.  For $\Lambda$CDM, $\alpha$ needs to be calculated 
numerically, although the limits of $k\tau \ll 1$ and
$k\tau \gg 1$ in matter domination can be derived analytically \cite{Schmidt14}.  
$\alpha(k,\tau)$ scales as $(k\tau)^2$ for $(k\tau) \ll 1$, i.e. for superhorizon tensor modes.  In the opposite limit $k\tau \to \infty$, $\alpha$ asymptotes to $2/5$ in matter domination.  

\begin{figure}
\includegraphics[angle=-90,width=0.4\textwidth]{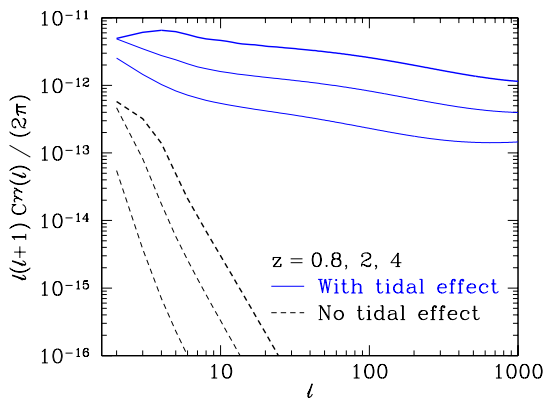}
\caption{The $B$-mode auto-power spectrum of galaxy shapes at $z=\{0.8,2,4\}$, from bottom to top. The blue lines show the contribution of alignments and lensing, while the black lines show the contribution of lensing alone. Alignments due to primordial tensor modes dominate the power spectrum.}
\label{fig:cgg}
\end{figure}

While this matching should only be seen as an order of magnitude estimate, Ref.~\cite{Schmidt14} showed that the tidal alignment effect dominates the shape correlation induced by tensor modes (in agreement with the previous estimate of \cite{Schmidt12}). Consequently we will neglect the contribution from gravitational lensing in this paper. Note that this is conservative, since (for $C_I > 0$) the alignment and lensing effects are positively correlated.  Keeping the above caveats in mind, we will assume the value of $C_I$ observed for scalar tidal fields for Luminous Red Galaxies up to $z\simeq 0.5$, $C_I = 0.12$ \citep{Blazek11}\footnote{Note that $C_I = C_1 \rho_{\rm crit}$ in the notation of \cite{Blazek11}, where $\rho_{\rm crit}$ is the critical density of the Universe today.}. We extrapolate this value to $z=2$, but the details of the evolution of $C_I$ with redshift are poorly understood. This coefficient also depends on the specific sample of galaxies considered, in particular, it depends on luminosity \cite{Hirata07}.   Figure \ref{fig:cgg} shows the auto-power spectrum of intrinsic alignment $B$-modes due to tensor modes as function of angular scale and for redshifts of $z=\{0.8,2,4\}$. The alignment contribution clearly dominates over the lensing at those redshifts.

%%%%%%%%%%%%%%%%%%%%%%%%%%%%%%%%%
%%%%%%%%%%%%%%%%%%%%%%%%%%%%%%%%%

The filter function for the CMB polarization is given by Eq.~(29) of \cite{Zaldarriaga/Seljak}, adjusted to our convention, 
\ba
F_l^{P}(k) \equiv\:& \frac1{\sqrt2}\int_0^{\tau_0} d\tau\: S_P^{(T)}(k,\tau) \left[2j_l'(x) + 4 \frac{j_l(x)}{x}\right]_{x=k\tau}\,,
\label{eq:Flpol}
\ea
where $S_P^{(T)}(k,\tau)$ is the $B$-mode polarization transfer function for tensors, given by \cite{Zaldarriaga/Seljak}
\be
S_P^{(T)}(k,\tau) = g\left(\frac{4\Psi}{x}+\frac{2\dot{\Psi}}{k}\right)+2\dot{g}\frac{\Psi}{x}
\ee
where $g(\tau)=\dot{\kappa}\exp(-\kappa)$ is the visibility function, $\kappa(\tau)$ is the optical depth, $x=k(\tau_0-\tau)$, the dots represent derivatives with respect to conformal time and $\Psi$ is a linear combination of the temperature and polarization perturbations due to gravitational waves, given by Eq.~(23) in \cite{Zaldarriaga/Seljak}. There are two contributions to the integral in \refeq{Flpol}, given by the two peaks of the visibility function: one at recombination ($z=1089$), and another around reionization. We perform the calculation of the angular power spectra \refeq{Cl} using a modified version of CAMB \cite{camb}.  The reionization history is parameterized by a hyperbolic tangent function of conformal time\footnote{http://cosmologist.info/notes/CAMB.pdf}. The reionization redshift, at which the ionization fraction is $0.5$, is assumed to be $z_{\rm re}=10.9$.

\begin{figure}
\includegraphics[width=0.45\textwidth]{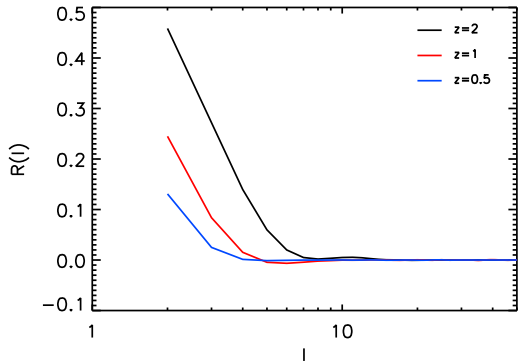}
\caption{Cross-correlation coefficient of shear and CMB B-modes as a function of
multipole moment and at redshifts $z=0.5$ (blue), $z=1$ (red) and $z=2$ (black).}
\label{fig:Rl}
\end{figure}

We define the cross-correlation coefficient of shear and polarization through
\be
C^{\g P}(l) = R(l) \sqrt{ C^{\g\g}(l) C^{PP}(l)}\,.
\ee
In Figure \ref{fig:Rl}, we show $R(l)$ as a function of scale and redshift. We only consider the cross-correlation up to $z=2$, given the expected limit of weak lensing surveys within this decade.  The cross-correlation relies on modes that affect both the CMB and galaxy shapes, and is dominated by the reionization contribution to the CMB B-mode polarization.  Furthermore, only very large-scale modes contribute to the cross-correlation because of the significant separation on the light cone between the reionization surface and the source galaxies at $z \leq 2$.  For these reasons, the cross-correlation is largest for the very lowest multipoles.  Thus, only the very largest angular scales, $l < 10$, are of interest for the cross-correlation.  Not surprisingly, the cross-correlation also grows with increasing galaxy redshift.  Note that we find a positive $R(l)$ whereas \citet{Dodelson10} found the opposite sign.  This issue, which is not relevant for our conclusions, is most likely due to a relative sign in the expression for the rotation $\omega$ (see App.~C of \cite{Schmidt12};  note that the lensing and intrinsic alignment effects of tensor modes are positively correlated for $C_I > 0$).

%--------------------------------------
%--------------------------------------

%%%%%%%%%%%%%%%%%%%%%%%%%%%%%%%%%%%%%%%%%%%%%%%%%%%%%%%%%%%%%%%%%%%%%%%%%%%
%%%%%%%%%%%%%%%%%%%%%%%%%%%%%%%%%%%%%%%%%%%%%%%%%%%%%%%%%%%%%%%%%%%%%%%%%%%
\section{Confirming the CMB signal with shapes}
\label{sec:sn1}

Consider overlapping CMB polarization and galaxy imaging surveys which cover a fraction $\fsky$ of the sky.  Each have noise: in case of the CMB, this is due to the instrument and foregrounds; for galaxy shapes, this is due to intrinsic galaxy ellipticities as well as B-modes due to instrument and shape measurement systematics and second order scalar contributions.  The latter are induced by lensing beyond the Born approximation, by the weighting of the shear field with the galaxy density, and by nonlinear tidal alignments.  For these source redshifts, the second contribution is expected to dominate (see also \cite{Schmidt12}).  In the following, we will neglect the foregrounds for the CMB, and non-primordial shear B-modes. The second order shear B-modes are only important for $l \gtrsim 10$ \cite{Schmidt12}, and we assume that B-modes due to systematics can be sufficiently mitigated using the measured shapes of stars\footnote{Weak lensing surveys usually rely on stars in our Galaxy to characterize the point spread function of the optical system used for the shear measurements. This characterization allows for the removal of spurious shear signals. See, for example, \cite{Huff14}.}. We emphasize that all these contributions are not expected to contribute to the cross-correlation between shear and CMB polarization and only modify the noise level assumed below.  B-modes in the CMB polarization due to lensing by density perturbations, which do correlate with the shear, are negligible compared to the primordial signal on the angular scales of interest ($l < 10$). 

Assuming Gausian noise, the variance of the angular cross power spectrum is then given by
\ba
{\rm Var}[C^{\gamma P}(l)] =\:& \left[\fsky (2l+1)\right]^{-1} \bigg\{ \vs
& \left[ C^{PP}(l) + N^P(l) \right] 
\left[ C^{\gamma\gamma}(l) + N^\gamma(l) \right]\vs
&  + [C^{\gamma P}(l)]^2\bigg\}\,,
\label{eq:VarCl}
\ea
where $N^X$ denote the noise contributions. $C^{\gamma P}(l)$ is 
distributed according to a $\chi^2$ distribution with $2l+1$ degrees 
of freedom, which approaches a Gaussian in the high-$l$ limit.  For
simplicity, we will approximate the distribution as a Gaussian for all $l$ in the
following, since our aim is an order-of-magnitude forecast.
The likelihood function for the measured cross-correlation $\hat C^{\g P}(l)$ is then given by
\be
\L\left( \{\hat C^{\g P}(l)\}_{l=2}^{l_{\rm max}} \right) \propto \exp \left[-\frac12
\sum_{l=2}^{l_{\rm max}} \frac{\left(\hat C^{\g P}(l) - \bar C^{\g P}(l)\right)^2}{{\rm Var}[\bar C^{\g P}(l)]}\right]\,,
\label{eq:likelihood}
\ee
where $\bar C^{\g P}(l)$ is the predicted cross-correlation and ${\rm Var}[\bar C^{\g P}(l)]$ is the variance shown in \refeq{VarCl} evaluated for the fiducial model.

We consider a fiducial CMB experiment with a noise level of $N^P=1$ $\mu K$~arcmin and a beam size of $\theta_{\rm FWHM}=1$ arcmin, which could be achieved with a future experiment such as the proposed CMB-Stage 4 \cite{Wu14}. We also consider a weak lensing survey where galaxy shapes are measured at  $z=2$.  The shear noise $N^\gamma = \sigma_\gamma^2 / 2n_g$ is determined by the galaxy density, for which we assume  $n_g=10$ red galaxies$/$arcmin$^2$, and the shape noise from the intrinsic ellipticities of galaxies, which we assume to be $\sigma_\gamma=0.3$.  All noise contributions are assumed to be $l$-independent on the scales of interest.  We study the case of $\fsky=1$.

First, how significantly can we rule out the hypothesis of $r=0$ using the cross-correlation, \emph{assuming that the CMB B-modes
are primordial ?}  The corresponding log likelihood ratio is given by 
\ba
-2\,\Delta\ln\L 
=\:& \sum_{l=2}^{l_{\rm max}} \fsky (2l+1) \frac{\left(C^{\g P}_{\rm signal}(l)\right)^2}
{ N^P(l) N^\gamma(l) }
\label{eq:SNq2}
\,.
\ea
For our fiducial experiment and a signal of $r=0.2$, this yields  $\sqrt{-2 \Delta\ln L} \simeq33$.  This result shows that a cross-correlation of primordial origin could be detected with high significance.  However, this significance is dominated by the CMB polarization, which on its own provides a much higher detection significance for the fiducial survey.  

The question we would really like to answer is: to what significance can we rule out a scenario that produces a signal in the CMB, $C^{PP}(l) > 0$, but no signal in the shapes, by using the cross-correlation? The relevant examples would be foregrounds or instrument systematics in the CMB polarization.
In this case, we calculate the likelihood of measuring a signal $C^{\g P}_{\rm signal}(l)$ given the hypothesis $\bar C^{\g P} = \bar C^{\g\g} = 0$, which yields a log likelihood ratio
\ba
-2\,\Delta\ln\L 
=\:& \sum_{l=2}^{l_{\rm max}} \fsky (2l+1) \frac{\left(C^{\g P}_{\rm signal}(l)\right)^2}
{ \left[ C^{PP}(l) + N^P(l) \right] N^\gamma(l) }
\label{eq:SNq1}
\,.
\ea
Evaluating this for a signal $C^{\g P}_{\rm signal}(l)$ and $C^{PP}(l)$ corresponding to $r=0.2$, our fiducial experiments yield $\sqrt{-2 \Delta\ln\L} \simeq 0.14$.

%%%%%%%%%%%%%%%%%%%%%%%%%%%%%%%%%%%%%%%%%%%%
\begin{figure}%[!ht]
\centering
\subfigure[$\,$Detection significance as a function of $n_g$.]{
\includegraphics[width=0.45\textwidth]{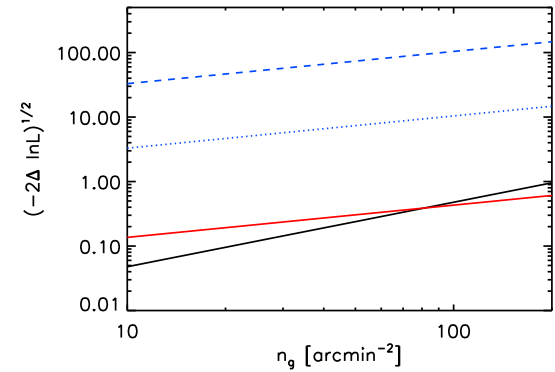}
\label{fig:sn1}
}
\subfigure[$\,$Detection significance as a function of $\sigma_\gamma$.]{
\includegraphics[width=0.45\textwidth]{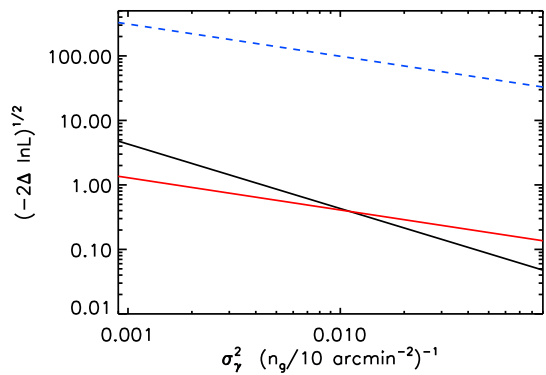}
\label{fig:sn2}
}
\subfigure[$\,$Same as panel (a), but for $r=0.02$.]{
\includegraphics[width=0.45\textwidth]{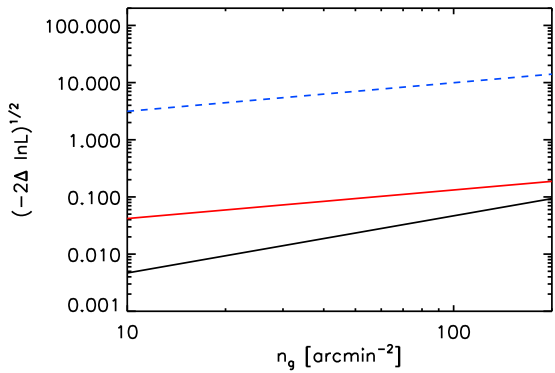}
\label{fig:sn3}
}
\caption{The significance for ruling out foregrounds (red curve, Eq. \ref{eq:SNq1}) and for ruling out $r=0$ (dashed blue curve, Eq. \ref{eq:SNq2}). This is presented as a function of the projected number density of red galaxies in the survey for $r=0.2$ (top panel) and $r=0.02$ (bottom panel). We also present it as a function of shape noise, $\sigma_\gamma^2$ for a number density of 10 galaxies per arcmin$^2$ and $r=0.2$ in the middle panel. The black curve shows the corresponding significance for the detection of $C^{\gamma\gamma}(l)$. The blue dotted curve in the top panel shows the decrease in the significance for ruling out $r=0$ when the CMB noise is increased by a factor of $10$.}
\label{fig:StoN_vs_shape_noise}
\end{figure}
%%%%%%%%%%%%%%%%%%%%%%%%%%%%%%%%%%%%%%%%%%%%%

We summarize our results in the three panels of Fig.~\ref{fig:StoN_vs_shape_noise}. Apart from the fiducial $r$, our results mainly depend on the shape noise in the shear survey.  In Fig.~\ref{fig:sn1}, we show the significance for detecting $r=0.2$ (dashed blue, Eq.~\ref{eq:SNq2}) along with the significance for ruling out foregrounds using the shear-CMB cross-correlation (red line, Eq.~\ref{eq:SNq1}) and shear auto-correlation (black line) as a function of the surface density of red galaxies with measured shapes in the survey. Fig.~\ref{fig:sn2} presents the same curves of Fig.~\ref{fig:sn1} but as a function of the dispersion in the intrinsic ellipticities, $\sigma_\gamma^2$, assuming $n_g=10$ red galaxies arcmin$^{-2}$. Finally, Fig.~\ref{fig:sn3} is similar to Fig.~\ref{fig:sn1} but for $r=0.02$. 
The impact of an increased noise in the CMB polarization measurements is shown in the dotted blue curve in Fig.~\ref{fig:sn1}, where we have modified our forecasts by increasing $N^P$ by a factor of 10. As expected, the significance for detecting the cross-correlation and the shear auto-correlation remain the same because the shape noise dominates over the CMB noise for the cases considered, but the significance for ruling out $r=0$ is proportional to the detector noise and hence it is reduced by a factor of 10 in this case. We have always assumed $f_{\rm sky}=1$, but the significance is expected to decrease proportionally to $\sqrt{f_{\rm sky}}$. 

For comparison, \emph{Euclid} will observe $15,000$ deg$^2$ ($f_{\rm sky}\sim$0.4) down to $24.5$ AB mag in the visible, with a resulting total $n_g=30$ galaxies per arcmin$^{2}$ and a median redshift of $0.9$ \cite{Laureijs11}. \emph{LSST}\footnote{http://www.lsst.org/lsst/} will cover $20,000$ deg$^2$ ($f_{\rm sky}\sim$0.5) down to $25.3$ AB mag in $i$-band, obtaining a total of $40$ galaxies per arcmin$^{2}$ with a median redshift of $1.2$ \cite{LSST08}. \emph{WFIRST-AFTA}\footnote{http://wfirst.gsfc.nasa.gov/} is better suited for our proposed study, expected to observe a total of $70$ galaxies per arcmin$^2$ between redshifts $0<z<2$, but it will only cover $2000$ deg$^2$ \cite{Spergel13}. The fraction of red galaxies is uncertain at high redshifts; we have assumed a constant fraction of $0.3$ but redshift dependence is expected \cite{Joachimi11}. 

Fig.~\ref{fig:StoN_vs_shape_noise} shows that the detection significance
for the shear auto correlation depends more strongly on the shape noise and can yield
a larger significance than the cross-correlation for small noise values. In order to elucidate this, we define the signal-to-noise value of the polarization auto-correlation for the hypothesis $r=0$ through
\be
\left(\frac{S}{N}\right)^P_{l} = [\fsky (2l+1)]^{1/2} \frac{C^{PP}_{\rm signal}(l)}{N^P(l)}\,,
\ee
and analogously for the shear auto-correlation $(S/N)^\g_l$.  Let us assume that there is a high $S/N$ detection in the polarization $(S/N)^P_l \gg 1$ on the scales of interest, so that $N^P$ is negligible compared to $C^{PP}$.  The log-likelihood ratio \refeq{SNq1} is then given by
\be
-2\,\Delta\ln\L 
= \sum_{l=2}^{l_{\rm max}} \fsky (2l+1) R^2(l) \frac{C^{\g\g}_{\rm signal}(l)}{ N^\gamma(l) }\,.
\ee
We can phrase this in terms of the $S/N$ for the shear auto-correlation:
\be
-2\,\Delta\ln\L
= \sum_{l=2}^{l_{\rm max}} [\fsky (2l+1)]^{1/2} R^2(l) \left(\frac{S}{N}\right)^\g_{l}\,.
\ee
Thus, given that $R(l)$ is only significant for $l < 10$, the signal-to-noise per $l$ of the detection of the cross-correlation can only be order 1 if the $S/N$ for the shear auto power spectrum is order 1, as confirmed by Fig.~\ref{fig:StoN_vs_shape_noise}.  
It is clear then that the $S/N$ for the cross-correlation will generically be smaller than the corresponding $S/N$ for the shear auto-correlation when either is required to be at least order $1$. The shear auto-correlation is thus expected to provide a confirmation of the CMB signal before the cross-correlation (in the sense of our second question, Eq. \ref{eq:SNq1}).  We stress again though that the cross-correlation is most likely the cleanest measurement regarding systematics and foregrounds.

%%%%%%%%%%%%%%%%%%%%%%%%%%%%%%%%%%%%%%%%%%%%%%%%%%%%%%%%%%%%%%%%%%%%%%%%%%%
%%%%%%%%%%%%%%%%%%%%%%%%%%%%%%%%%%%%%%%%%%%%%%%%%%%%%%%%%%%%%%%%%%%%%%%%%%%
\section{Discussion}
\label{sec:concl}

In this paper, we have considered the possibility of using the cross-correlation between shear and CMB B-mode polarization to confirm or rule out the detection of gravitational waves from inflation.  Due to the entirely independent foregrounds and systematics of both measurements, the cross-correlation is an exceptionally clean test of the primordial origin of the B-mode polarization. In comparison to previous work on the cross-correlation between CMB B-mode polarization and weak lensing shear \cite{Dodelson10}, we have included the contribution to shear from the tidal alignments of galaxies, which have been shown to be the dominant tensor mode contribution to galaxy shape correlations \cite{Schmidt12,Schmidt14}.

We have found that full-sky overlapping CMB and galaxy surveys, with noise levels currently expected for a \emph{Euclid}-like survey, will most likely not be able to confirm a primordial tensor mode background at the $r=0.2$ level. This is due to the shape noise in the shear survey. 
We have not explored the optimal weighting of the alignment signal with redshift and alignment strength. Instead, we have assumed a fixed redshift of $z=2$ for all galaxies with shapes, which is optimistic for an \emph{Euclid}-like survey. Given current uncertainty in the dependence of $C_I$ on galaxy color and luminosity, we have assumed a fixed value for $C_I$ consistent with current measurements of alignments of red galaxies.  
However, our results could change towards the positive if galaxy populations with alignment strength significantly larger than the value assumed here can be identified: the detection significance in cross and auto-correlation scales as $C_I$ and $C_I^2$, respectively. 
The dispersion in galaxy ellipticities is astrophysical in origin and is thus not expected to decrease simply by reducing the instrumental noise in shear surveys.  However,
new techniques for significantly reducing the shape noise \cite{Huff13} could improve the prospects.  Finally, next-generation surveys beyond \emph{Euclid}, 
\emph{WFIRST-AFTA}, and \emph{LSST} will likely yield a further significant reduction in the noise by providing an even larger number of measured galaxy shapes.
While the measurements proposed here are clearly extremely challenging, we stress that other avenues for complementary constraints on tensor modes, such as 21cm observations or direct detection of gravitational waves through satellite interferometry are likely to be even more distant in the future. The enormous ramifications for fundamental physics of a primordial gravitational wave background however clearly justify the significant efforts needed to provide independent confirmation.

%%%%%%%%%%%%%%%%%%%%%%%%%%%%%%%%%%%%%%%%%%%%%%%%%%%%%%%%%%%%%%%%%%%%%%%%%%%
%%%%%%%%%%%%%%%%%%%%%%%%%%%%%%%%%%%%%%%%%%%%%%%%%%%%%%%%%%%%%%%%%%%%%%%%%%%
\acknowledgments

We would like to thank David Spergel, Michael Strauss and Scott Dodelson for their comments on this manuscript.  
CD was supported by the National Science Foundation grant number AST-0807444, NSF grant number PHY-088855425, and the Raymond and Beverly Sackler Funds.

%\newpage
\bibliography{cds_alignmentsCMB_070114}

\end{document}